\documentstyle [12pt] {article}

\newcommand{\req}[1]{(\ref{#1})}
\newcommand{\sect}[1]{ \section{#1} \setcounter{equation}{0} }

\newcommand{\bam}[1]{\left( \begin{array}{#1}}
\newcommand{\ab}{\end{array} \right)}

\newcommand{\ve}{\left( \begin{array}{r}}
\newcommand{\ev}{\end{array} \right)}
\newcommand{\ar}{\left( \begin{array}{rr}}
\newcommand{\ra}{\end{array} \right)}
\newcommand{\arr}{\left( \begin{array}{rrrr}}
\newcommand{\arrr}{\left( \begin{array}{rrrrrr}}

\newcommand{\eqr}{\begin{eqnarray}}
\newcommand{\rqe}{\end{eqnarray}}
\newcommand{\eq}{\begin{equation}}
\newcommand{\qe}{\end{equation}}
\newcommand{\mreq}[1]{\req{#1:start} -- \req{#1:end}}

\newcommand{\oned}{{\bf 1}_d}
\newcommand{\half}{\mbox{$\frac{1}{2}$}}

\newcommand{\ziii}{$Z_3$}

\newcommand{\zvii}{$Z_7$}

\newcommand{\es}{\mbox{$ e^{\ast} $}}

\newcommand{\qs}{\mbox{$ Q^{\ast} $}}
\newcommand{\qt}{\mbox{$ Q^T $}}

\newcommand{\om}{\mbox{$ \Omega $}}

\newcommand{\omt}{\mbox{$ \Omega ^T $}}
\newcommand{\omi}{\mbox{$ \Omega ^{-1} $}}

\newcommand{\h}{\mbox{$\chi$}}

\newcommand{\sltz}{\mbox{$SL(2;{\bf Z})$}}
\newcommand{\sq}{{\cal R}}
\newcommand{\sqt}{\sq^T}
\newcommand{\ig}{\mbox{$ g^{-1}$}}

\newcommand{\tgroup}{\mbox{$ {\cal G}_{\cal N} $}}
\newcommand{\ogroup}{\mbox{$ {\cal G}_{\cal O} $}}
\newcommand{\gmo}{\mbox{$ \tilde{\cal G}_{\cal O} $}}

\def\KK{{\rm I\kern -.2em  K}}
\def\NN{{\rm I\kern -.16em N}}
\def\RR{{\rm I\kern -.2em  R}}
\def\ZZ{{\small{\rm Z}\kern -.34em Z}}
\def\ZZZ{{\small{\rm Z}\kern -.5em Z}}
\def\QQ{{\rm \kern .25em
             \vrule height1.4ex depth-.12ex width.06em\kern-.31em Q}}
\def\CC{{\rm \kern .25em
             \vrule height1.4ex depth-.12ex width.06em\kern-.31em C}}

\newcommand{\csb}{complex spacetime basis}
\newcommand{\at}{a^T}

\newcommand{\be}{\begin{equation}}
\newcommand{\ee}{\end{equation}}
\newcommand{\ba}{\begin{array}}
\newcommand{\ea}{\end{array}}
\newcommand{\bea}{\begin{eqnarray}}
\newcommand{\eea}{\end{eqnarray}}
\newcommand{\halb}{{1\over 2}}
\newcommand{\viert}{{1\over 4}}
\newcommand{\dritt}{{1\over 3}}
\newcommand{\sext}{{1\over 6}}
\newcommand{\vier}{\\ [4 pt]}
\newcommand{\acht}{\\ [8 pt]}
\newcommand{\zwoelf}{\\ [12 pt]}

\newcommand{\alp}{\alpha^\prime}
\newcommand{\raw}{\rightarrow}
\newcommand{\th}{\theta}

\newcommand{\hp}{\hspace{25pt}}
\newcommand{\wa}{W^\ast}
\newcommand{\wi}{W^{-1}}
\newcommand{\lra}{\leftrightarrow}

\newcommand{\bc}{\begin{center}}
\newcommand{\ec}{\end{center}}

\begin{document}
\begin{titlepage}
\title{
{\normalsize August 1992} \hfill {\normalsize MPI-Ph/92-61}\\
\hfill {\normalsize TUM-TH-147/92}\\
\hfill\\
Modular Groups for Twisted Narain Models\thanks{Supported by
Deutsche Forschungsgemeinschaft}}
\vspace{5.cm}
\author{
Jens Erler \\
{\small Max-Planck-Institut f\"ur Physik} \\
{\small --- Werner-Heisenberg-Institut ---} \\
{\small F\"ohringer Ring 6} \\ {\small Postfach 40 12 12} \\
{\small W -- 8000 Munich 40 (Germany)}
\\ \\ and \\ \\
	Micha\l\ Spali\'nski \\
{\small Instytut Fizyki Teoretycznej}\\
{\small Uniwersytetu Warszawskiego}\\
{\small Warszawa (Polska)}}
\date{\quad}
\maketitle
\begin{abstract}
\noindent
We demonstrate how to find modular discrete
symmetry groups for $Z_N$ orbifolds.
The $Z_7$ orbifold is treated in detail as a non-trivial example of
a $(2,2)$ orbifold model. We give the generators of the modular group for
this case which, surprisingly, does not contain $\sltz^3$ as had been
speculated. The treatment models with discrete Wilson
lines is also discussed. We consider examples which demonstrate that discrete
Wilson lines affect the modular
group in a non-trivial manner. In particular, we show that it is possible for
a Wilson line to break $SL(2,{\bf Z})$.
\end{abstract}
\thispagestyle{empty}
\end{titlepage}

\setcounter{page}{1}

\sect{Introduction}

One of the main tasks of the string theorist today is to
find the true vacuum configuration of the heterotic string.
On one hand one should try to phenomenologically constrain
possible vacua and on the other hand it is clearly important to
understand the nature of the vacuum degeneracy and the dynamical effects
which could lead to the determination of the ground state.

The space of string vacua is locally parametrized by moduli,
which are marginal deformations of the underlying conformal field
theory~\cite{DVV}.
An intriguing feature of string compactifications is that the
natural parametrizations of the moduli space label the points in a
redundant way, since all physical quantities are invariant under the action
of some discrete group acting on the moduli.

In the low energy effective theory the moduli correspond to vacuum expectation
values of massless scalar fields which have flat potentials to all
orders in perturbation theory~\cite{DS,S}.
It would thus be of considerable phenomenological importance, if the
degeneracy of the vacua described by these moduli could be lifted.
One attempt~\cite{FLST,FLT,FT} in this direction
is based on the assumption that the modular discrete symmetry group
derived in the framework of string perturbation theory remains
unbroken after taking into account non-perturbative effects.

It is also interesting, that the modular group typically includes
duality symmetries~\cite{SS,SW,GRV,DHS}, which relate small and large radii.
Like all modular
symmetries, it must be used to restrict the moduli space to a
fundamental domain. Duality is a stringy property, not found in a
point particle context. If signs of such a symmetry would survive down to
low energies, they would constitute an indication of a compactified string
theory at some level.

A simple yet quite rich class of compactifications which allow a detailed
study of the moduli space are $Z_N$ orbifolds~\cite{DHVW,IMNQ}, also
referred to as
twisted Narain compactifications~\cite{N,NSW}. These models are
parametrized by
constant background fields which  are interpreted as the
orbifold moduli. Having the moduli explicitly in hand makes it possible
to answer detailed questions about the structure of moduli space and
in particular the discrete symmetries~\cite{MS1,MS2,EJN,EJNS}. This is
especially exciting
since this class of models has very promising prospects for
phenomenology.

Until recently the only examples where the modular group had been
identified were tori and symmetric orbifolds in one and two
dimensions~\cite{DVV,SS,SW,GRV,LLW,LMN2}.
In this paper we investigate two compactifications, each serving as a
representative of a large class of models. First we consider the
$Z_7$ orbifold falling into the class of vacua with $(2,2)$ world
sheet supersymmetry. Motivated by the results in~\cite{LLW,LMN2}, where
the modular group \sltz\ was found, it was sometimes conjectured that this
group is independently realized for any complex modulus which appears, so
that the modular group would always be \sltz\ to some power. This
conjecture became known as the ``maximal discrete symmetry hypothesis''.
This belief suffered a setback with the discovery that the modular group
for the complex structure modulus of the mirror quintic
was different~\cite{COPG}. Nevertheless, it was still speculated that the
conjecture could be true in orbifold compactifications.
For the $Z_7$ orbifold the symmetry group expected on the basis of the
maximal discrete symmetry hypothesis would be $SL(2,{\bf Z})^3$.
This is in fact not the case - instead we find a surprisingly rich
structure, where $SL(2,{\bf Z})$
subgroups are not realized independently on each of the individual complex
moduli associated with the three complex planes distinguished by the twist.

A related longstanding problem concerned the case where quantized Wilson lines
are present. These models are of particular phenomenological
interest, as three generation models of $SU(3) \times SU(2) \times
U(1)^n$ can be constructed in this class~\cite{INQ}.
The question whether a duality symmetry in moduli
space can be defined has been questioned seriously, since it was
realized~\cite{EJN}
that the duality transformation rules from the torus applied to these
cases lead to asymmetric orbifolds. Here we will show, for the $Z_3$ orbifold,
that turning on discrete Wilson lines in fact breaks \sltz\ to a
subgroup.

Section \ref{nara} reviews the Narain model and its modular symmetries.
Section \ref{twinamo} discusses symmetric $Z_N$ orbifolds of the Narain
model,
stressing various points which are important for further developments.
It also discusses the modular symmetries of these models.
Our main results are presented in sections \ref{z7} and
\ref{wl}, where the modular groups of the \zvii\ orbifold and the \ziii\
orbifold with a Wilson line background are described. Section \ref{z7} also
contains a direct proof of the fact that \sltz\ is the modular
group for the \ziii\ orbifold in two compact dimensions.

\sect{The Narain model}
\label{nara}

The Hilbert space of a Narain model~\cite{N,NSW} is built by applying the
standard
mode creation operators to states of given momentum and winding. The states
may therefore be labelled by the left and right momenta,
associated with the holomorphic and antiholomorphic sectors of the
world sheet CFT and by the mode numbers. The entire dependence on the
moduli lies in the momenta labelling the ground state of a
tower of integer-spaced states.

Narain models are obtained by compactifying $d$ coordinates on a torus,
whose geometry is specified by an even, self-dual lattice with a
Lorentzian metric.
The set of such lattices is labelled by $d(d+16)$ continuous
parameters, which may be interpreted in terms of expectation values of
background fields: the lattice metric $g$, an antisymmetric
tensor field $b$ and Wilson lines $a$.

In our case the lattices in
$(16+d,d)$ dimensions are of relevance and can be described in terms
of the
spectra of the left and right momenta, which due to the compactification
become discrete. The parametrization in terms of constant background fields
reads\footnote{In
this and in the following two sections we set the Regge slope $\alp$ to 1.}
\be \ba{rclcccl}
P_L=(m + (g - h) n - a^T C l,& l+a n& )=(&p_L&,&\tilde{p}_L&),\label{PN} \acht
P_R=(m - (g+  h) n - a^T C l,&    0   & )=(&p_R&,&      0    &).
\ea \ee
These expressions are understood in the lattice basis, so that
the windings $n$, momenta $m$ and gauge quantum numbers $l$ are
all integer valued vectors. In this formulae $g$ denotes the lattice
metric (a $d\times d$ matrix) and
$a$ is the Wilson line (a $16\times d$ matrix). Furthermore we defined
\be
h=b+\half a^T C a,
\ee
where $b$ is the antisymmetric ``axionic'' background and $C$
denotes the $E_8 \times E_8$ Cartan metric.

Dropping moduli independent oscillator contributions, the conformal dimensions
are given by
\be \ba{l}
L_0 = \viert p_L g^{-1} p_L + \halb \tilde{p}_L C \tilde{p}_L , \acht
\bar{L}_0 = \viert p_R g^{-1} p_R .
\ea \ee
For many purposes it is convenient to consider the linear combinations,
\bea
H = L_0 + \bar{L}_0, \\
P = L_0 - \bar{L}_0,
\eea
which are interpreted as the world sheet energy and momentum, respectively.
Since both quantities are quadratic in the quantum numbers,
they can be written in the form
\eq \label{chi}
H = \half u^T \h u,
\qe
\eq
P = \half u^T \eta u,
\qe
where
\eq
u = \bam{ccc} n \\
             m \\
	     l \ab ,
\qe
\eq
\eta = \bam{ccc} 0 & {\bf 1}_d & 0 \\
	{\bf 1}_d & 0 & 0 \\
	0 & 0 & C \ab ,
\qe
and
\eq
\h = \bam{ccc}
	 (g+h^T) \ig (g+h) & - h^T \ig &  (g+h^T) \ig \at C \\
	             - \ig h   &    \ig &       -\ig \at C \\
	   C a \ig (h + g) & -Ca\ig & C+Ca\ig\at C
\ab .
\qe
Thus $u$ is a $2 d + 16$ component integer vector, while $\h$ and $\eta$
are $2 d + 16$ dimensional square matrices and ${\bf 1}_d$ denotes the
identity matrix in $d$ dimensions. The above
formulation is useful, since all the moduli dependence resides in $\h$,
while $\eta$ is constant over moduli space.

Discrete symmetries in moduli
space can be studied by looking for linear transformations of the quantum
numbers, which preserve the spectrum. Such transformations can be written
as
\eq \label{thetra}
u \longrightarrow S_{\om}(u) = \omi u.
\qe
The requirement that $P$ should be preserved by this transformation
translates to the following condition on the integer matrix $\Omega$:
\eq
\omt \eta \om = \eta.           \label{etacon}
\qe
The group formed by such matrices will be denoted by \tgroup.

To be a symmetry of the spectrum, the transformation \req{thetra} must
also leave $H$ invariant. Since $H$ depends on the moduli, invariance can
only be achieved if the background is appropriately transformed. The
required transformation law is
\eq
\h \longrightarrow S_{\om}(\h) = \omt \h \om.          \label{trans}
\qe
This defines the action of \tgroup\ on the moduli.

\sect{Twisted Narain models}
\label{twinamo}

The untwisted
sector of the orbifold Hilbert space is obtained by a projection of the
Hilbert space of a Narain compactified heterotic string and adding twisted
sectors needed to maintain modular invariance. The orbifold projection is
defined in terms of an automorphism group of the Narain
lattice~\cite{IMNQ,MS2,EK}. To construct a
$Z_N$ orbifold~\cite{DHVW} one may
proceed by choosing
an element $\sq$ of \tgroup\ (the defining matrix) which generates a
$Z_N$ group and
projecting onto invariant states.

The action of the twist on the quantum numbers is given by
\eq
u \longrightarrow u^\prime = \sq u.
\qe
For the Narain lattice to admit such an
automorphism group, the background has to satisfy the consistency
condition (see \req{chi})
\eq
\sqt \h \sq = \h .                     \label{chicon}
\qe
This is a set of homogenous linear equations determining the specific form
of the background
which admits a given twisting. The solutions to this equation are
parametrized by the moduli.

Requiring that the orbifold be left -- right symmetric restricts the
possible form of the defining matrix $\sq$. One must have
\eq
n \longrightarrow n^\prime = Q n,
\qe
where $Q$ is the twist matrix in the lattice basis~\cite{IMNQ}. One must
also require that the twist acts trivially on $\tilde{p}_L$, since the
corresponding (vanishing) components of $P_R$ are not affected by the
twist. It then follows that
\eq
l \longrightarrow l^\prime = l + a (\oned - Q) n.
\qe
This way one arrives at the conclusion that $\sq$ must have the form
\eq \label{rgenf}
\sq = \bam{lll}
	Q & 0 & 0 \\
	\alpha & \beta & \gamma \\
	a (\oned - Q) & 0 & \oned \ab ,
\qe
where $\alpha,
\beta, \gamma$ are matrices whose form will be
determined by the requirement that $\sq$ should belong to \tgroup.
The use of \req{etacon} leads to the result that
\eqr \label{etsol}
\alpha & = &  \qs \delta + \half \at C a (\oned-Q) + \half (\oned-\qs) \at
C a , \nonumber \\
\beta & = & \qs , \\
\gamma & = & (\oned-\qs) \at C , \nonumber
\rqe
where $\delta$ is an antisymmetric integer matrix, which is not determined
by the condition \req{etacon}. The $\ast$ denotes the inverse transpose of
a matrix, i.e. $A^\ast \equiv (A^T)^(-1)$.

Since $\sq$ has to be an integer matrix in order to have a well defined
meaning on the quantum numbers, it follows from the above that the
Wilson line has to satisfy
\eq
\qs \delta + \half \at C a (\oned-Q) + \half (\oned-\qs) \at C a \in {\bf Z},
\label{wlc:start}
\qe
\eq
a (\oned-Q) \in {\bf Z}.
\label{wlc:end}
\qe
These conditions were originally found by Iba\~nez et al.~\cite{IMNQ}.

One must also require that the defining matrix should preserve $\h$ in the
sense of \req{chicon}. This yields
\eq
\qt g Q = g,                                \label{concs1}
\qe
\eq
\qt b Q = b+\delta ,                             \label{concs2}
\qe
where $\delta$ is the antisymmetric integer matrix appearing in
\req{etsol}. These equations determine the backgrounds in terms of
independent deformation parameters (the moduli). Of course
\req{concs2} may not admit a solution with a non-vanishing $\delta$. This is
in fact the case in all the examples discussed in this paper. Note finally
that \req{chicon} imposes no conditions on the Wilson lines. There can
however be additional
conditions on the Wilson lines coming from modular invariance~\cite{DHVW}
(level matching).

Even though twisted Narain models are constructed directly from the
untwisted ones, not all the modular discrete symmetries of the orbifold
moduli
space are inherited by the orbifold. Only those symmetries survive, which
are compatible with the orbifold projection, as will now be reviewed.

The conditions for modular discrete symmetries in orbifold models were
studied  in~\cite{MS1,MS2,EJN}. The basic requirement to be imposed is
that symmetry  transformations map physical states at a reference
background  to physical states at the transformed
background. By physical states we mean here twist invariant states of the
model under consideration\footnote{Note that in~\cite{EJN} a more general
situation is considered, where the physical states of one model are mapped
to physical states of another yet equivalent model.}.
This is the case for those $\Omega \in {\cal G}_{\cal N}$, which preserve the
$Z_N$ group in the sense that
\eq
\om \sq \omi = \sq^p,                   \label{comc}
\qe
for some $p = 1, \dots , N-1$. This condition selects a subgroup
of \tgroup, which is the discrete symmetry group of the orbifold
compactification. This group will be referred to as \ogroup, defined as
\eq \label{ogroup}
\ogroup := \{\om \in \tgroup | \ \om \sq \omi = \sq^p,\ p \in {\bf Z} \}.
\qe

The condition \req{comc} may be called the compatibility condition for the
discrete transformation specified by $\om$. It guarantees that
transformations induced by elements of \ogroup\ are
modular symmetries of the orbifold compactification.

Not all the elements belonging to \ogroup\
induce non-trivial transformations in moduli space. For example
the defining matrix $\sq$ itself trivially satisfies \req{comc}
with $p=1$ and by construction leaves the background unchanged.
Thus, the representation of
\ogroup\ on the orbifold moduli space is not faithful. It is therefore
convenient to define an equivalence relation in \ogroup\ by declaring all
elements, which induce the same transformation of the moduli, to be
equivalent.
The group of discrete symmetries acting on the moduli space may then be
defined as the quotient of \ogroup\ by the above equivalence. To avoid
misunderstandings, this quotient group will be denoted by \gmo.

The canonical duality transformation for the Narain model is the
transformation induced by
\eq
\Omega = \bam{ccc}
	0 & {\bf 1}_d & 0 \\
	{\bf 1}_d & 0 & 0 \\
	0 & 0 & {\bf 1}_{16} \ab .
\qe
In the orbifold case the above transformation is
typically not modular. Instead, one has
to consider the following family of transformations~\cite{MS1,MS2},
\eq \label{wdu}
\Omega_S := \bam{ccc}
	0      & W^{-1} & 0 \\
	W^T & 0 & 0 \\
	0      & 0 & {\bf 1}_{16} \ab ,
\qe
where $W$ is an integer matrix with determinant $\pm 1$. In the case
without Wilson lines, these
transformations are modular whenever $W$ satisfies
\eq \label{wmat}
W Q = \qs W  ,
\qe
and they generate a subgroup of \ogroup. The action induced on the
backgrounds (via \req{trans}) is
\bea
g \longrightarrow W \frac{1}{g+b} g  \frac{1}{g-b} W^T , \\
b \longrightarrow - W \frac{1}{g+b} b  \frac{1}{g-b} W^T .
\eea
In the case when there is a
non-vanishing Wilson line this transformation is not modular, and no
similar general expression for a modular duality-type symmetry is known.

There is also a number of so called axionic shift symmetries, generated by
\eq
\Omega_T := \bam{ccc}
    {\bf 1}_d  &      0    &         0 \\
    \epsilon     & {\bf 1}_d &         0 \\
    0          &      0    & {\bf 1}_{16} \ab ,
\qe
where $\epsilon$ is an antisymmetric integer matrix. The corresponding action
on the background is $b\longrightarrow b+\epsilon$. Generically a number of
these symmetries
remain modular on the orbifold; in fact any axionic modulus may be shifted
by an integer.

Moduli spaces of supersymmetric compactifications posess a complex
structure, which means in particular that
one can define complex coordinates on moduli space in a natural way, and
many quantities will take a simple form in these variables. Already in the
case of two dimensional orbifolds~\cite{LLW,LMN2},
which posess two real moduli, it is
clearly of great advantage to express modular symmetries in terms
of a single complex modulus. In models with more
parameters this advantage becomes a prerequisite, if one is interested
in obtaining any useful formulae at all.

\sect{The duality group for the \zvii\ orbifold}
\label{z7}

In this section we discuss in detail the $Z_7$ orbifold, which can be
defined as a Coxeter twist~\cite{IMNQ} in the root lattice of $SU(7)$ and
deformations
thereof. The twist matrix in the lattice basis can be chosen as
\eq
Q = \arrr 0 & 0 & 0 & 0 & 0 & -1 \\
           1 & 0 & 0 & 0 & 0 & -1 \\
           0 & 1 & 0 & 0 & 0 & -1 \\
           0 & 0 & 1 & 0 & 0 & -1 \\
           0 & 0 & 0 & 1 & 0 & -1 \\
           0 & 0 & 0 & 0 & 1 & -1
    \ra,
\qe
and $\sq$ is given by \req{rgenf}, \req{etsol} with $\delta = 0$ and $a =
0$. The solutions to \req{concs1}--\req{concs2} read
\eq
g =  \arrr 2 g_0 & g_1 & g_2 &  g_3  & g_3 & g_2 \\
           g_1 & 2g_0 & g_1 & g_2 & g_3 & g_3 \\
           g_2 & g_1 & 2g_0 & g_1 & g_2 & g_3 \\
           g_3 & g_2 & g_1 & 2g_0 & g_1 & g_2 \\
           g_3 & g_3 & g_2 & g_1 & 2g_0 & g_1 \\
           g_2 & g_3 & g_3 & g_2 & g_1 & 2g_0
     \ra,   \label{g76}
\qe
with the constraint $ \sum\limits_{i=0}^{3} g_i = 0 $, and
\eq
b = \arrr  0  & b_1  & b_2  & b_3 & -b_3 & -b_2 \\
          -b_1 & 0  & b_1  & b_2 & b_3 & -b_3   \\
          -b_2 & -b_1 & 0  & b_1 & b_2 & b_3 \\
          -b_3 & -b_2 & -b_1 & 0  & b_1 & b_2 \\
          b_3 & -b_3 & -b_2 & -b_1 & 0  & b_1 \\
          b_2 & b_3 & -b_3 & -b_2 & -b_1 & 0
    \ra.
\qe
Here any three of the $g_i$ can be taken as the (real) metric moduli, and
the $b_i$ are the (real) axionic moduli.

In the \csb, where the complex moduli will be defined, the twist is simply
given by
\be
\th ={\rm diag}\;
(\alpha,\alpha^{-1},\alpha^2,\alpha^{-2},\alpha^3,\alpha^{-3}),
\ee
where $\alpha \equiv \exp(2\pi i/7)$.
The transformation matrix $e$, relating the two bases via
\be \label{edef}
       \th e = e Q
\ee
is given by
 \eq
e = \bam{llllll}
\alpha^2 & \alpha^3 & \alpha^4 & \alpha^5 & \alpha^6 & 1 \\
\alpha^5 & \alpha^4 & \alpha^3 & \alpha^2 & \alpha & 1 \\
\alpha^4 & \alpha^6 & \alpha & \alpha^3 & \alpha^5 & 1 \\
\alpha^3 & \alpha & \alpha^6 & \alpha^4 & \alpha^2 & 1 \\
\alpha^6 & \alpha^2 & \alpha^5 & \alpha & \alpha^4 & 1 \\
\alpha   & \alpha^5 & \alpha^2 & \alpha^6 & \alpha^3 & 1
\ab .
\qe

The background fields in the \csb\ are denoted by $G$ and $B$. They
are connected with their lattice basis counterparts by the
equations\footnote{When working with a complex
basis one has to use the hermitian conjugate instead of transposition. Thus
the star now denotes the inverse of the hermitian conjugate.}
\be
   G \pm B = e^\ast (g \pm b) e^{-1}.
\ee
{}From this one can read off the complex moduli. The result is
\eq \ba{ll} \label{commod}
t_1 = - i\tan({\pi \over 7}) &[g_1 +{s_1^2\over s_3^2}g_2 +{s_2^2\over
s_3^2}g_3]
    + b_1 + {s_2 \over s_1 } b_2 + {s_3 \over s_1 } b_3,       \acht

t_2 = - i\tan({2\pi\over 7}) &[g_1 +{s_2^2\over s_1^2}g_2 +{s_3^2\over
s_1^2}g_3]
    + b_1 - {s_3 \over s_2 } b_2 - {s_1 \over s_2 } b_3,       \acht

t_3 = - i\tan({3\pi\over 7}) &[g_1 +{s_3^2\over s_2^2}g_2 +{s_1^2\over
s_2^2}g_3]
    + b_1 - {s_1 \over s_3 } b_2 + {s_2 \over s_3 } b_3,
\ea \qe
where
\eq \ba{c}
  s_k \equiv \sin(2\pi k/7),  \vier
  c_k \equiv \cos(2\pi k/7).
\ea \qe
The formulae \req{commod} involve a choice of normalization, which will be
explained below. The requirement that the metric be positive definite here
translates into the statement that each complex modulus takes values
in the complex upper half-plane.

The generators of the axionic shifts $T_k$ are defined as shifts of
the real moduli $b_i$ leading trivially to modular symmetries. Thus one has
\eq \label{t1}
T_k(t_m) = t_m + a_{km},
\qe
with
\eq \label{t2}
a_{km} = \bam{ccc}
1 & 1 & 1 \\
s_2/s_1 & -s_3/s_2 & -s_1/s_3 \\
s_3/s_1 & -s_1/s_2 & s_2/s_3
\ab .
\qe
The first row in $a_{km}$ shows that one of the axionic shifts acts on the
complex moduli in the usual way. This in fact motivated the normalization
chosen in \req{commod}.

At this point is is interesting to check whether one can find
combinations of the axionic shifts which would be decoupled on the $t_k$,
as would be
necessary if $\sltz^3$ were to be realized here. Consider shifting the
$b_i$ by
$k_i$, where $k_i$ are integers. One would like to be able to find a
choice of these
integers such that under the above transformation $t_1$ would undergo a
unit shift,
but the remaining moduli would be invariant (and similarly for $t_2$ and
$t_3$). For this to happen one has to solve a system of three linear equations
which follow from \req{commod} and the above requirements. This system of
equations has no integer solutions. This shows, that on the complex moduli
associated with the complex planes
it is not possible to define an $ \sltz^3 $.

Having found the overall shift symmetry $T_1$ we have to answer the
question, whether there is a related duality transformation $S_1$
transforming all $t_i$ into their negative inverses and so completing
an overall $SL(2,{\bf Z})$. This is indeed the case and the
corresponding $W$ matrix satisfying condition \req{wmat} is
\eq
W_1 =
\arrr
0 & -1 & 0 & 0 & 0 & 0 \\
1 & 0 & -1 & 0 & 0 & 0 \\
0 & 1 & 0 & -1 & 0 & 0 \\
0 & 0 & 1 & 0 & -1 & 0 \\
0 & 0 & 0 & 1 & 0 & -1 \\
0 & 0 & 0 & 0 & 1 & 0
\ra .
\qe
It will be shown below that it is also possible to find symmetry
transformations
$S_2$ and $S_3$ serving as the $SL(2,{\bf Z})$ ``partners'' of
$T_2$ and $T_3$, i.e.
\be \label{sl}
   S_k^2 = 1,   \hspace{2cm}  (S_k T_k)^3 = 1.
\ee
It is important to emphasize, however, that the transformations $S_k$
do not mutually commute. This can be checked
by studying the explicit transformation rules on the complex
moduli as given later on.

The condition \req{comc} for group elements $\Omega \in$ \ogroup\
involves the discrete parameter $p$. The
transformations described above were all solutions with $p=1$ and they
were all realized holomorphically on the moduli. The solutions with
$p \neq 1$ lead to transformations
which are non-holomorphic~\cite{E}. In these cases the matrices
$\Omega$, although commuting with the projector onto twist invariant
states, do not
commute with the defining matrix $\sq$. For our $Z_7$ case
we could have chosen equivalently any power $Q^n$ with
$n \neq 0\; {\rm mod} \; 7$ to be the defining twist and solutions
with non-trivial $p$ just correspond to exchanging different twist powers.
As for the moduli, they are in general permuted and
complex conjugated. If one chooses such a transformation $R$ for $p=3$ or
$p=5$, then one can generate all the other ones with non-trivial $p$ by
combining $R$ with the transformations of the holomorphic subgroup.

In the present case we find $R$ in the form
\eq
\Omega_R := \bam{ccc}
    V          &      0      &         0 \\
    0          &      V^\ast &         0 \\
    0          &      0      & {\bf 1}_{16} \ab ,
\qe
with
\eq
V =
\arrr
0 & 0 & 0 & 1 & 0 & -1 \\
0 & 1 & 0 & 0 & 0 & -1 \\
0 & 0 & 0 & 0 & 0 & -1 \\
0 & 0 & 0 & 0 & 1 & -1 \\
0 & 0 & 1 & 0 & 0 & -1 \\
1 & 0 & 0 & 0 & 0 & -1
\ra ,
\qe
satisfying
\be
V Q = Q^3 V.
\ee
The resulting transformation on the moduli reads
\be
   R: \; (t_1,t_2,t_3) \raw (-2 c_1 \bar{t_2}, -2 c_2 t_3, -2 c_3 t_1).
\ee
The essential property of $R$ is that
\eq \ba{c}
R T_1 R^{-1} = T_3, \vier
R T_3 R^{-1} = T_2,
\ea \qe
i.e., by conjugating $T_1$ with $R$ one obtains the other axionic shift
symmetries. Observe also that
\be
   R^3: \; (t_1,t_2,t_3) \raw (-\bar{t_1},-\bar{t_2},-\bar{t_3}) .
\ee
Thus $R^3$ gives rise to a charge conjugation symmetry.

It is clear that the transformations
\be \ba{c}
   S_2 := R^2   S_1 R^{-2}, \vier
   S_3 := R S_1 R^{-1},
\ea \ee
satisfy the relations \req{sl}. The transformations $S_k$ act on the
moduli according to
\eq
S_k (t_m) = p_{km}/t_m ,
\qe
where the numbers $p_{km}$ are given by
\eq
(p_{km}) =
\bam{ccc}
-1 & -1 & -1 \\
-4 c_1^2 & -4 c_2^2 & -4 c_3^2 \\
{-1\over 4 c_3^2} & {-1\over 4 c_1^2} & {-1\over 4 c_2^2}
\ab .
\qe
{}From this it is easy to see that different $S$ transformations and thus
the correspon\-ding \sltz\ groups do not commute. Note, that the composition of
two different duality transformations yields a simultaneous rescaling
of all moduli, e.g.
\be
   S_1 S_2: \; (t_1,t_2,t_3) \raw (\frac{t_1}{4 c_1^2},
\frac{t_2}{4 c_2^2}, \frac{t_3}{4 c_3^2}),
\ee
and that this is an infinite order symmetry.
By composing such elements with $S_1, T_1$ one can construct an infinite
number of non-commuting \sltz\ groups.

In summary, we found three generators $S:=S_1$, $T:=T_1$ and $R$
satisfying a number of relations. The basic ones are given by
\be \ba{c}
   S^2=(ST)^3=R^6=1 , \vier
   (TR)^6=(TR^3)^2=(SR^3)^2=1 , \vier
   (SRTR^{-1})^7=1 .
\ea \ee
These relations define a group which does not contain $SL(2,{\bf Z})^3$.

Given an arbitrary string of transformations generated
by $S$, $T$ and $R$ one can always permute all factors of $R$ to one
side obtaining a product of a power of $R$ with a purely holomorphic
transformation. For a classification we can thus concentrate on the
holomorphic ($p=1$) symmetry transformations. The price to be paid is,
however, that the number of holomorphic generators is larger: one has to
take all six $S_i, T_i$.

We now proceed to characterize matrices $\Omega$ which belong to \gmo.
Recall that such matrices must satisfy two conditions: the invariance
condition \req{etacon}, and the compatibility relation \req{comc},
which we consider for $p=1$. The latter is a linear equation for $\Omega$
and can readily be solved to give a form of $\Omega$ in terms of a number of
integer parameters. This can then be substituted into \req{etacon}, which
gives a set of quadratic equations for the integer parameters.
Some of
these equations can be solved, while further parameters may be eliminated
by observing that certain $\Omega$'s act trivially on the moduli.
One is then left with a number of quadratic relations between the remaining
integers.

To see
the mechanics of this procedure let us first apply it to the example of the
\ziii\ orbifold in two compact dimensions (without Wilson lines), where the
group \gmo\ is known to be \sltz\ ~\cite{LLW}. The twist in the
lattice basis can be taken in the form
\eq \label{z3q}
Q = \bam{rr}
0 &  -1 \\
1 & -1
\ab ,
\qe
and $\sq$ is given by \req{rgenf}, \req{etsol} with $\delta = 0$ and $a = 0$.
The twist matrix in the complex basis reads
\eq
\theta ={\rm diag}\;
(\alpha,\alpha^{-1}),
\qe
where now $\alpha=\exp(2\pi i/3)$. Since there is no Wilson line in this
example, the gauge lattice will be dropped in the formulae given below.

The solution to \req{comc} can be written in the form
\eq \label{omc}
\Omega =
\arr
u_3 - u_5 & -u_5 & u_6 & -u_1 - u_6 \\
u_5 & u_3 & u_1 & u_6 \\
u_7 & u_4 + u_7 & u_2 + u_8 & -u_8 \\
-u_4 & u_7 & u_8 & u_2
\ra ,
\qe
depending on eight integers $u_i$. To reduce this further it is
convenient to pass to the complex basis. As before, the basis transformation
matrix $e$ is defined by \req{edef}. In this case it reads
\eq
e = \bam{ll}
\alpha^2 & 1 \\
\alpha & 1
\ra .
\qe
It is easy to see that $\Omega$ must be transformed according to
\eq
\tilde{\Omega} = E^\ast \Omega E^T,
\qe
where $E$ is the block diagonal matrix
\eq
E = \arr
e & 0 \\
0 & \es
\ra .
\qe
One finds
\eq \label{omtex:start}
\tilde{\Omega} =
\bam{llll}
z_1 & 0 & z_2  & 0 \\
0 &  \bar{z}_1 & 0 & \bar{z}_2 \\
z_3 & 0 & z_4 & 0 \\
0 & \bar{z}_3 & 0 & \bar{z}_4
\ab ,
\qe
where
\eq
\ba{ll} \label{omtex:end}
z_1 &= u_3 + \alpha u_5 , \\
z_2 &= \frac{1}{3} ((1 + 2 \alpha)u_1 + (2  + \alpha)u_6) , \\
z_3 &= -(1+ 2 \alpha) u_4 + (1 - \alpha) u_7 , \\
z_4 &= u_2 - \alpha^2 u_8 .
\ea
\qe
It is useful to look at this from the following point of view:
write
\eq \label{omtil}
\tilde{\Omega} =
\ar a & b \\
    c & d
\ra ,
\qe
where $a,b,c,d$ are $2\times 2$ blocks of the form
\be \label{theblocks}
x = r_x \bam{cc}
\exp{i \phi_x} & 0 \\
0 & \exp{- i \phi_x} \ab .
\ee
The quantities $r_x$ and $\phi_x$ will be referred to as the radius and the
phase of
a block. It turns out that the phases can be determined, as will now be
shown.

The condition \req{etacon} in the complex basis implies
\bea
a^\dagger c + c^\dagger a = 0, \\ \label{etc:start}
b^\dagger d + d^\dagger b = 0, \\
a^\dagger d + c^\dagger b = 1.     \label{etc:end}
\eea
{}From \mreq{etc} it follows that
\bea
\phi_c = \phi_b, \\
\phi_a = \phi_d,
\eea
and
\eq
\phi_a - \phi_c = \pm \frac{\pi}{2} .
\qe
This way three of the four phases are fixed. The remaining one is fixed
by observing that multiplying $\tilde{\Omega}$ by a matrix of the form
\eq
\Lambda = {\rm diag} (\exp{i\psi}, \exp{- i\psi},\exp{i\psi},\exp{-i\psi})
\qe
for any $\psi$ does not change the transformation induced by
$\tilde{\Omega}$ on the moduli. This
allows us to set one of the phases (say $\phi_a$) equal to zero. Thus all
phases have been found
by using the equation \req{etacon} and eliminating the redundancy in
$\Omega$. Only the four radii remain.

The conditions on the blocks in \req{omtil} can be
stated in the form
\eq \label{recon} \ba{c}
Re( b) = Re (c) = 0, \vier
Im (a) = Im (d) = 0.
\ea \qe
The arguments given above can be used to reduce the number of integer
parameters in \req{omc}. One simply enforces the equations \req{recon} on
$\tilde{\Omega}$. Looking at \mreq{omtex} one sees that $u_i$ $(i=5,\dots,8)$
must be set to zero, so only four integer parameters
remain. Back in the lattice basis $\Omega$ reads
\eq \label{fom}
\Omega =
\arr
u_3 & 0 & 0 & -u_1 \\
0 & u_3 & u_1 & 0 \\
0 & u_4 & u_2 & 0 \\
-u_4 & 0 & 0 & u_2
\ra ,
\qe
and the constraint \req{etacon} gives
\eq \label{detcon}
u_1 u_4 - u_2 u_3 = 1,
\qe
which is the standard determinant condition of \sltz.

Using \req{z3q} in \req{concs1}--\req{concs2} gives the backgrounds
\eq \label{bk3:start}
g = R^2
\bam{rr}
2 & -1 \\
-1 & 2
\ab,
\qe
\eq \label{bk3:end}
b = \bam{rr}
0 & x \\
-x & 0
\ab ,
\qe
in terms of real moduli $R^2$ and $x$.
Passing to the complex basis one finds the complex modulus
\eq
\lambda = - i \sqrt{3} R^2 + x
\qe
One may now
check, that the $\Omega$ given in \req{fom} induces the standard action of
\sltz\ on the complex modulus $\lambda$:
\eq
\lambda \longrightarrow \lambda^\prime =
	\frac{u_1 + \lambda u_2}{u_3 + \lambda u_4} .
\qe

The arguments given above are easily generalized to the $Z_7$ orbifold.
Solutions to \req{comc} depend in this case on $24$ integers.
The matrix $\tilde{\Omega}$ now has the form
\be
\tilde{\Omega} =
\bam{cccccc}
   a_1 & 0 & 0 & b_1 & 0 & 0 \\
   0 & a_2 & 0 & 0 & b_2 & 0 \\
   0 & 0 & a_3 & 0 & 0 & b_3 \\
   c_1 & 0 & 0 & d_1 & 0 & 0 \\
   0 & c_2 & 0 & 0 & d_2 & 0 \\
   0 & 0 & c_3 & 0 & 0 & d_3 \ab ,
\ee
where $a_i, b_i, c_i, d_i$ are $2\times 2$ blocks of the form \req{theblocks}.
We thus have 12 radii and 12 phases.

The condition \req{etacon} for $\tilde{\Omega}$ implies
\bea
  a_k^\dagger c_k + c_k^\dagger a_k = 0, \\
  b_k^\dagger d_k + d_k^\dagger b_k = 0, \\
  a_k^\dagger d_k + c_k^\dagger b_k = 1.
\eea
It follows from this, that
\be \ba{c}
Arg(a_k) = Arg(d_k), \vier
Arg(b_k) = Arg(c_k),
\ea \ee
and
\eq
Arg(a_k) = Arg(c_k) \pm \pi/2 ,
\qe
where $Arg(x)$ denotes the phase of the block $x$.
The above equations determine $9$ of the $12$ phases in terms of $3$,
which can however be chosen arbitrarily, using the freedom to
multiply $\tilde{\Omega}$ by the following matrix, which induces a trivial
transformation on the moduli:
\eq
\Lambda = {\rm diag}(\lambda_1,\lambda_2,\lambda_3,
                     \lambda_1,\lambda_2,\lambda_3 ).
\qe
Here the $\lambda_i$ are diagonal $2\times 2$ blocks
\eq
\lambda_k = {\rm diag} (\exp{i\psi_k}, \exp{-i\psi_k}),
\qe
and the $\psi_k$ are arbitrary.

Thus one can choose $a_k, d_k$ to be real,
and $b_k, c_k$ to be purely imaginary. Imposing this in $\tilde{\Omega}$
leaves $12$ integers, which must satisfy quadratic constraints following
{}From \req{etacon}, in analogy with \req{detcon}.

The result of the calculation described above is as follows. The matrix
$\Omega$ depends on $12$ integers denoted by $u_i$, and is given
in the appendix. The quadratic constraints read
\bea
u_{5} u_{7} + u_{4} u_{8} + u_{6} u_{8} - u_{4} u_{9} - u_{1} u_{10} -
u_{1} u_{11} + u_{3} u_{11} - u_{2} u_{12} = 0, \\
u_{6} u_{7} + u_{5} u_{8} + u_{4} u_{9} - u_{6} u_{9} - u_{2} u_{10} +
u_{3} u_{10} - u_{1} u_{11} - u_{1} u_{12} + u_{2} u_{12} = 0, \\
u_{4} u_{7} - u_{4} u_{8} + u_{5} u_{8} - u_{5} u_{9} + u_{6} u_{9} -
u_{1} u_{10} + u_{2} u_{10} - u_{2} u_{11} + u_{3} u_{12} = 1.
\eea
It is possible to show, that the transformations induced by such matrices
transform the complex moduli according to
\eq
t_k \longrightarrow {t_k}^\prime = \frac{a_k + b_k t_k}{c_k + d_k t_k} ,
\qe
where the parameters $a_k, b_k, c_k, d_k$ can be expressed in terms of the
integers $u_i$.

\section{Quantized Wilson lines and duality}
\label{wl}

In this section we will come to the question, whether non-trivial
(quantized) Wilson lines can change the results obtained so far.
At first sight it might appear that these Wilson lines essentially
affect only the gauge lattice, since on one hand they break the
gauge group
while on the other hand they do not seem to change the twist in the
space part which stays in its symmetric form. However, as explained
in section~\ref{nara}, this is not quite the case, since quantized
Wilson lines are not just usual background parameters like the moduli.
Recall, that the corresponding defining matrix mixes up
all types
of quantum numbers. Indeed, in~\cite{EJN} it was shown, that naive
generalizations of duality lead to the class of asymmetric orbifolds,
thus not representing modular symmetries.
Consequently, it is interesting to ask whether one can
define some kind of duality and, motivated from the results of
the preceding section, what role is played by the group $SL(2,{\bf Z})$
in these cases.

In order to illustrate what happens, we discuss two simple
examples~\cite{E}. Thereby we will for simplicity assume
that the shift vector
({\em gauge embedding of the point group}) acts in one $E_8$ and the
Wilson line ({\em gauge embedding of the space group}) is connected
to the other one. Then we can disregard the former one as well as the
world sheet superfermions which also play a passive role for our
discussion.

\begin {figure}[h]
\begin {picture}(160,50)(-70,0)
  \put (35,5){\circle{10}} \put (32,2){\mbox{\small{1}}}
  \put (40,5){\line(1,0){15}}
  \put (60,5){\circle{10}} \put (57,2){\mbox{\small{2}}}
  \put (65,5){\line(1,0){15}}
  \put (85,5){\circle{10}} \put (82,2){\mbox{\small{3}}}
  \put (90,5){\line(1,0){15}}
  \put (110,5){\circle{10}} \put (107,2){\mbox{\small{4}}}
  \put (115,5){\line(1,0){15}}
  \put (135,5){\circle{10}} \put (132,2){\mbox{\small{5}}}
  \put (140,5){\line(1,0){15}}
  \put (160,5){\circle{10}} \put (157,2){\mbox{\small{6}}}
  \put (165,5){\line(1,0){15}}
  \put (185,5){\circle{10}} \put (182,2){\mbox{\small{7}}}
  \put (135,10){\line(0,1){15}}
  \put (135,30){\circle{10}} \put (132,27){\mbox{\small{8}}}
\end {picture}
\caption {Dynkin diagram of $E_8$ \label{Dynkin}}
\end {figure}
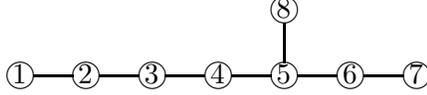

We start with the one-dimensional case often referred to as the
orbicircle, where the twist action is given by negation of
winding and momentum. If we assign the gauge quantum numbers according
to figure~\ref{Dynkin} and use the
Wilson line\footnote{A non-trivial Wilson line with only one entry would
not satisfy $a^2 = a^TCa \in {\bf Z}$ necessary for level matching.}
\eq
   a=\halb (0,0,0,0,0,0,1,1),
\qe
then we find the right moving Narain momentum
\eq\label{QRA2}
   P_R=\halb m-{1\over 2\alp}r^2n + \viert (l_5 + l_6 - n) - \halb (l_7 + l_8),
\qe
where we reinserted the dimensionful parameter $\alp$. Notice the factor
of $1/4$ in this expression, which has no counterpart in the case
of vanishing Wilson line. Motivated by this we rewrite (\ref{QRA2}) as
\eq
   P_R=-{2\over\alp}r^2[\viert n-{{\alp}\over 8r^2}(2m-n+l_5+l_6-2l_7-2l_8)].
\qe
{}From dimensional considerations we conclude that the expression
in parenthesis gives the new winding. Recalling that scalar products
are built with the inverse metric we find invariance of $P_R^2$
when transforming
\eq\label{DUALA2} \ba{rcl}
   r &\raw& {\alp\over 2r}, \acht
   n &\lra& 2m - n + l_5 + l_6 - 2l_7 - 2l_8 = 2m - n - 2 a^T C l.
\ea \qe
The remaining task is to extend this to a complete $O(9,1;{\bf Z})$
transformation, which can be done with
\eq\label{DUALA3} \ba{rcl}
   m &\raw& -m, \acht
   l &\raw&  l - 2 a l.
\ea \qe
This shows that there is indeed a $Z_2$ symmetry in moduli space, which
can be interpreted as a generalized duality transformation. Note, however,
that the radius is no longer transformed like in the case with vanishing
$a$, where $r \raw {\alp\over r}$. We see that the self-dual radius has
changed. This is in perfect agreement with the results in~\cite{EJN},
where a simple model with a discrete antisymmetric background was
presented and anticipated that these kinds of quantized background fields
in fact stand on the same footing.

We will now discuss the example of the $Z_3$-orbifold, where
in case of vanishing Wilson lines the modular group is known~\cite{DFF}
to be $SU(3,3;{\bf Z})$. This model admits up to three {\em independent}
Wilson lines~\cite{INQ} and in using two ones one can construct models with
three generations of $SU(3) \times SU(2) \times U(1)^n$~\cite{IKNQ}.
Here, again for simplicity, we will restrict ourselves to one
independent Wilson line,
\eq
   a_1=a_2={1\over 3} (1,0,0,0,0,0,1,1),
\qe
where already the important new features occur. This means that we deal
effectively with a two-dimensional $Z_3$-orbifold, where the modular group
for $a=0$ is $PSL(2,{\bf Z})$~\cite{LMN2}.

The twist matrix in this case is given by \req{z3q}
and we obtain the right moving Narain momentum
\eq\label{QRA3} \ba{rcl}
   {P_R}_i &=& \halb m_i - {1\over 2\alp} (g+b)_{ij} n^j +
   \sext (l_2 + l_5 + l_6 - n^1 - n^2)- \dritt (l_1 + l_7 + l_8) \acht
           &=& \halb m_i - {1\over 2\alp} (g+b)_{ij} n^j + \sext t_i,
\ea \qe
where we introduced the vector $t$ with ($a_1=a_2$) equal components
$t_1=t_2 \in {\bf Z}$. In analogy to the previous example we rewrite
this to
\eq
   P_R= -{3\over \alp} (g+b) W^T [{1\over 6} \wa n -
         \wa {\alp \over 18(g+b)} \wi (3Wm + Wt)],
\qe
where as in section~\ref{nara} we introduced a matrix $W$.
Consequently we try the following ansatz for duality:
\eqr
   \label{gbhat3}   \hat{g}+\hat{b} &=& \wa {{\alp}^2 \over 9(g+b)} \wi, \vier
   \label{nhat3}            \hat{n} &=& W (3m + t), \vier
   \label{mhat3} 3\hat{m} + \hat{t} &=& \wa n, \vier
                          \hat{P}_R &=& -\wa {\alp \over 3(g+b)} ,
\rqe
where we chose to indicate the transformed quantities by hats.
Notice the factor of 3 multiplying the momenta on the right hand side of
(\ref{nhat3}). Thus to make sure that in the transformed theory the quantum
number $\hat{n}^1+\hat{n}^2$ runs over all integers we have to require
that the sum of matrix elements in $W$ does not vanish mod 3. The solutions
to equation (\ref{wmat}) are given by
\eq\label{WDEF} \ba{ccc}
   W_1=\left( \ba{rr} 0 & 1 \\ -1 & 0 \ea \right), \hp
&  W_2=\left( \ba{rr} 1 & 0 \\  1 & 1 \ea \right), \hp
&  W_3=\left( \ba{rr} 1 & 1 \\  0 & 1 \ea \right), \zwoelf
   W_4=\left( \ba{rr} 1 & 0 \\  0 &-1 \ea \right), \hp
&  W_5=\left( \ba{rr} 0 & 1 \\  1 & 1 \ea \right), \hp
&  W_6=\left( \ba{rr} 1 & 1 \\  1 & 0 \ea \right),
\ea \qe
as well as there negatives and in fact {\em none\/} of them meets our
requirement. This shows that discrete Wilson lines can break duality
symmetry~\cite{E}. An immediate conclusion is that the group
$PSL(2,{\bf Z})$ is broken as well, a fact with far reaching
consequences. This group is well known in the mathematical literature since
a long time~\cite{KF} and this knowledge has been used intensively
for the construction of four dimensional low energy field theories~\cite{FLST}.
For cases with quantized Wilson lines one obviously has to consider
different groups. So let us look for the unbroken symmetry group.

We will try to modify the ansatz (\ref{nhat3}) and in order to circumvent
the above problem we add a {\em partial axionic shift\/}, i.e.\ we shift
the momentum quantum number by a non-integer amount, which by itself is
not a symmetry:
\eq
   \hat{n} = W (3m + t + \epsilon_{k/3} n),
\qe
with
\eq\label{PARTDELTA}
   \epsilon_{k/3} = \left( \ba{cc} 0 & k \vier -k & 0 \ea \right), \hp
              k \in {\bf Z} \not\in 3{\bf Z}.
\qe
But (\ref{mhat3}) shows that the $m$ transformation is problematic as well,
since it follows
\eq\label{MDIF}
   3(\hat{m}_1-\hat{m}_2)=(\wa_{11}-\wa_{21})n^1+(\wa_{12}-\wa_{22})n^2,
\qe
so that the right hand side must be a multiple of three for the ansatz
to work. Again this cannot be achieved for any $W$ and we modify the
ansatz further to
\eq
   3\hat{m} + \hat{t} + \hat{\epsilon}_{\hat{k}/3} \hat{n} = \wa n,
\qe
where as in (\ref{PARTDELTA}) we defined $\hat{\epsilon}_{\hat{k}/3}$
with elements $\hat{k}$. Then (\ref{MDIF}) takes on the form
\eq
   3(\hat{m}_1-\hat{m}_2)=(\wa_{11}-\wa_{21}-\hat{k})n^1+
                          (\wa_{12}-\wa_{22}-\hat{k})n^2.
\qe
Now one can indeed achieve\footnote{See reference~\cite{E} for more
details.} that the r.h.s.\ is a multiple of 3 and the simplest solution arises
for $W_1$,
\eq\ba{rcl}
   n &\raw& U n + 3 W_1 m - 3 \sigma_3 a^T C l, \acht
   m &\raw& U^\ast m, \acht
   l &\raw& l + 3 a \sigma_3 m,
\ea \qe
with $\sigma_3$ the Pauli matrix and
\eq
   U=\left( \ba{rr} 0 & -1 \\ 1 & 2 \ea \right).
\qe
The above construction involves an inversion $\sigma$ and partial shifts
$\tau$. We normalize the complex modulus\footnote{The quantities $g$ and
$x$ are defined in \mreq{bk3}.}
\eq
   \lambda := {3\over \alp}(x + i\sqrt{\det g})
\qe
in such a way that we have
\eq \ba{rl}
   \sigma: \hp \lambda &  \raw -{1\over \lambda}, \acht
   \tau:   \hp \lambda &  \raw \lambda + 1.
\ea \qe
This notation shows that the modular group at hand is the subgroup
of $PSL(2,{\bf Z})$ generated by
\eq\label{wlgroup} \ba{l}
   A := \tau \sigma \tau, \acht
   B := \tau^3,
\ea \qe
where $B$ refers to the usual axionic shift symmetry which is
clearly also present. In matrix notation wie can represent the
action of $A$ as
\eq
   \hat{g}+\hat{b}-{\alp \over 3}\epsilon_{1/3} =
           \wa {{\alp}^2 \over 9(g+b+{\alp \over 3}\epsilon_{1/3})}\wi ,
\qe
and the complex modulus transforms as
\eq\label{TST}
   \lambda \raw {\lambda \over \lambda +1}.
\qe
With the help of (\ref{wlgroup}) one easily finds the relation
\eq
   (AB)^3=1,
\qe
whereas $A$ no longer squares to one as would be necessary for
$PSL(2,{\bf Z})$.
It can be shown~\cite{E} that for the ordinary $Z_3$ orbifold
there exists another symmetry generator $C$ with a non-holomorphic
action,
\eq
   C: \hp \lambda   \raw - \bar\lambda,
\qe
giving rise to an exchange of particles with antiparticles in the
twisted sector. It also survives in the Wilson line case and with the help
of
\eq
   \tau C \tau = C
\qe
one readily proves the relations~\cite{E}
\eq\label{DURCHTAUSCH}
   C^2=(AC)^2=(BC)^2=1.
\qe
Note, that none of the order two operations above can be utilized to
define a new kind of duality transformation, since all of them
posess fixed curves instead of fixed points as is the case for duality.
Similarly, no $PSL(2,{\bf Z})$ subgroup can reappear. Notice also, that
the relations (\ref{DURCHTAUSCH}) can be used to commute all generators
$C$ to one side of a given string of operations showing that $C$
corresponds to a $Z_2$ factor commuting with the holomorphic part of
the modular group.

Having established the breakdown of duality in this example, we are
confronted with the question whether there is still an operation
(1) possessing fixed points and (2) mapping small radii to large ones.
Whereas $A$ leaves fixed only the origin, which does not belong to the
upper half-plane, the combination $AB$ does in fact
the job. It has a fixed point at $\sqrt{3} e^{i \pi/6}$ and property
(2) is obvious from the construction.

\vskip 0.5cm
{\bf
\begin{center}
Acknowledgements
\end{center}
}
\vskip 0.2cm

We would like to thank Hans-Peter Nilles for many enlightening discussions
on this and related topics. We also thank Stefan Theisen for helpful
comments. A large part of this work was done while M.S. was in M\"unchen as
a fellow of the Alexander von Humboldt-Stiftung.

\appendix

\sect{Appendix}

For typographical reasons it is convenient to present the matrix $\Omega$
discussed in section \ref{z7} in block form:
$$
\Omega =
\ar
A & B \\
C & D
\ra
$$
where the blocks are given by
$$
A =
\arrr
u_{3} &
0 &
u_{2} &
u_{1} &
u_{1} &
u_{2}
 \\
-u_{2} &
-u_{2} + u_{3} &
-u_{2} &
0 &
u_{1} - u_{2} &
u_{1} - u_{2}
 \\
-u_{1} + u_{2} &
-u_{1} &
-u_{1} + u_{3} &
-u_{1} &
-u_{1} + u_{2} &
0
 \\
0 &
-u_{1} + u_{2} &
-u_{1} &
-u_{1} + u_{3} &
-u_{1} &
-u_{1} + u_{2}
 \\
u_{1} - u_{2} &
u_{1} - u_{2} &
0 &
-u_{2} &
-u_{2} + u_{3} &
-u_{2}
 \\
u_{2} &
u_{1} &
u_{1} &
u_{2} &
0 &
u_{3}
\ra
$$ \nonumber
$$
B =
\arrr
0 &
-u_{6} &
-u_{5} &
-u_{4} &
u_{4} &
u_{5}
 \\
u_{6} &
0 &
-u_{6} &
-u_{5} &
-u_{4} &
u_{4}
 \\
u_{5} &
u_{6} &
0 &
-u_{6} &
-u_{5} &
-u_{4}
 \\
u_{4} &
u_{5} &
u_{6} &
0 &
-u_{6} &
-u_{5}
 \\
-u_{4} &
u_{4} &
u_{5} &
u_{6} &
0 &
-u_{6}
 \\
-u_{5} &
-u_{4} &
u_{4} &
u_{5} &
u_{6} &
0
\ra \nonumber
$$
$$
C=
\arrr
0 &
-u_{9} &
-u_{8} &
-u_{7} &
-u_{8} &
-u_{9}
 \\
u_{9} &
0 &
-u_{8} &
-u_{7} - u_{8} + u_{9} &
-u_{7} - u_{8} + u_{9} &
-u_{8}
 \\
u_{8} &
u_{8} &
0 &
-u_{7} &
-u_{7} - u_{8} + u_{9} &
-u_{7}
 \\
u_{7} &
u_{7} + u_{8} - u_{9} &
u_{7} &
0 &
-u_{8} &
-u_{8}
 \\
u_{8} &
u_{7} + u_{8} - u_{9} &
u_{7} + u_{8} - u_{9} &
u_{8} &
0 &
-u_{9}
 \\
u_{9} &
u_{8} &
u_{7} &
u_{8} &
u_{9} &
0
\ra \nonumber
$$

$$
D =
\arrr
u_{12} &
u_{11} &
u_{10} &
0 &
-u_{10} &
-u_{11}
 \\
0 &
u_{11} + u_{12} &
u_{10} + u_{11} &
u_{10} &
-u_{10} &
-u_{10} - u_{11}
 \\
-u_{11} &
u_{11} &
u_{10} + u_{11} + u_{12} &
u_{10} + u_{11} &
0 &
-u_{10} - u_{11}
 \\
-u_{10} - u_{11} &
0 &
u_{10} + u_{11} &
u_{10} + u_{11} + u_{12} &
u_{11} &
-u_{11}
 \\
-u_{10} - u_{11} &
-u_{10} &
u_{10} &
u_{10} + u_{11} &
u_{11} + u_{12} &
0
 \\
-u_{11} &
-u_{10} &
0 &
u_{10} &
u_{11} &
u_{12}
\ra \nonumber
$$

\newpage

\newcommand{\bi}[1]{\bibitem{#1}}

\newcommand{\np}{\mbox{\em {Nucl. Phys.} {\bf B}}}
\newcommand{\pl}{\mbox{\em {Phys. Lett.} {\bf }}}
\newcommand{\cmp}{\mbox{\em {Comm. Math. Phys. }}}
\newcommand{\prd}{\mbox{\em {Phys. Rev.} {\bf D}}}
\newcommand{\npb}[3]{\mbox{{\em {Nucl. Phys.}} {\bf B {#1}} (19{#2}) {#3} }}

\end{document}